\documentclass[aps,preprint,floats,superscriptaddress,showpacs]{revtex4}

\usepackage{graphicx}

\begin{document}

\title{Consequences of the local spin self-energy approximation on
the heavy Fermion quantum phase transition}

\author{Ping Sun}
\author{Gabriel Kotliar}
\affiliation{
Center for Materials Theory,
Department of Physics and Astronomy,
Rutgers University,
Piscataway, NJ 08854-8019}
\date{\today}

\begin{abstract}
We show, using the periodic Anderson model, that the local spin
self-energy approximation, as implemented in the extended dynamical
mean field theory (EDMFT), results in a first order phase transition
which persists to $T=0$. Around the transition, there is a finite
coexistence region of the paramagnetic and antiferromagnetic (AFM)
phases. The region is bounded by two critical transition lines which 
differ by an electron-hole bubble at the AFM ordering wave vector.
\end{abstract}

\pacs{71.27.+a, 71.10.Hf,72.15.Qm,75.20.Hr }

\maketitle

\section{Introduction.}
\label{sec-01}

Competing Kondo and RKKY interactions in Heavy Fermion materials
induce a quantum phase transition \cite{doniach,varma} near which
various deviations from the Landau-Fermi liquid behavior are observed
experimentally \cite{stewart}. Among the well-studied heavy Fermion
compounds is $CeCu_{6-x}Au_x$ \cite{lohneysen} on which neutron
scattering and magnetometry experiments showed \cite{schroder} that,
in the quantum critical region, the spin susceptibilities, both the
homogeneous one and that at the antiferromagnetic (AFM) ordering wave
vector, followed,

\begin{equation}
\label{eq-01}
   \chi^{-1}(\vec{q},T)=\left[T^{\alpha}+\theta^{\alpha}(\vec{q})\right]/C
\end{equation}

\noindent with $T$ the temperature, $\theta(\vec{q})$ a momentum
dependent function which is a measure of the distance from the
critical wave-vector, and $C$ the Curie constant. In the experiments
it was found the exponent $\alpha \sim 0.75$, unlike $\alpha=1$ in the
standard Curie-Weiss law. This same behavior was found, within
experimental error, to be followed by the neutron scattering data
taken at the other wave vectors. The disentanglement of the
temperature and momentum dependences in the inverse spin
susceptibilities led to the suggestion \cite{si0} that the self-energy
of the spin-spin interaction be local in space and correspond to the
frequency-dependent part of the observed $\chi^{-1}$. The theoretical
formulation of this observation turned out to be the extended
dynamical mean field theory (EDMFT).\\

The EDMFT is a method developed to study, within the local self-energy
approximation, correlated electron systems in the existence of
non-local interactions \cite{edmft,chitra}, which, in the context of
heavy Fermions, is the RKKY interaction. It allows the dynamical
screening of the bare interactions. As a result, EDMFT is able to
describe the competing RKKY and Kondo interactions in a more balanced
way than the original DMFT.\\

\begin{figure}[ht]
\includegraphics[width=7.5cm]{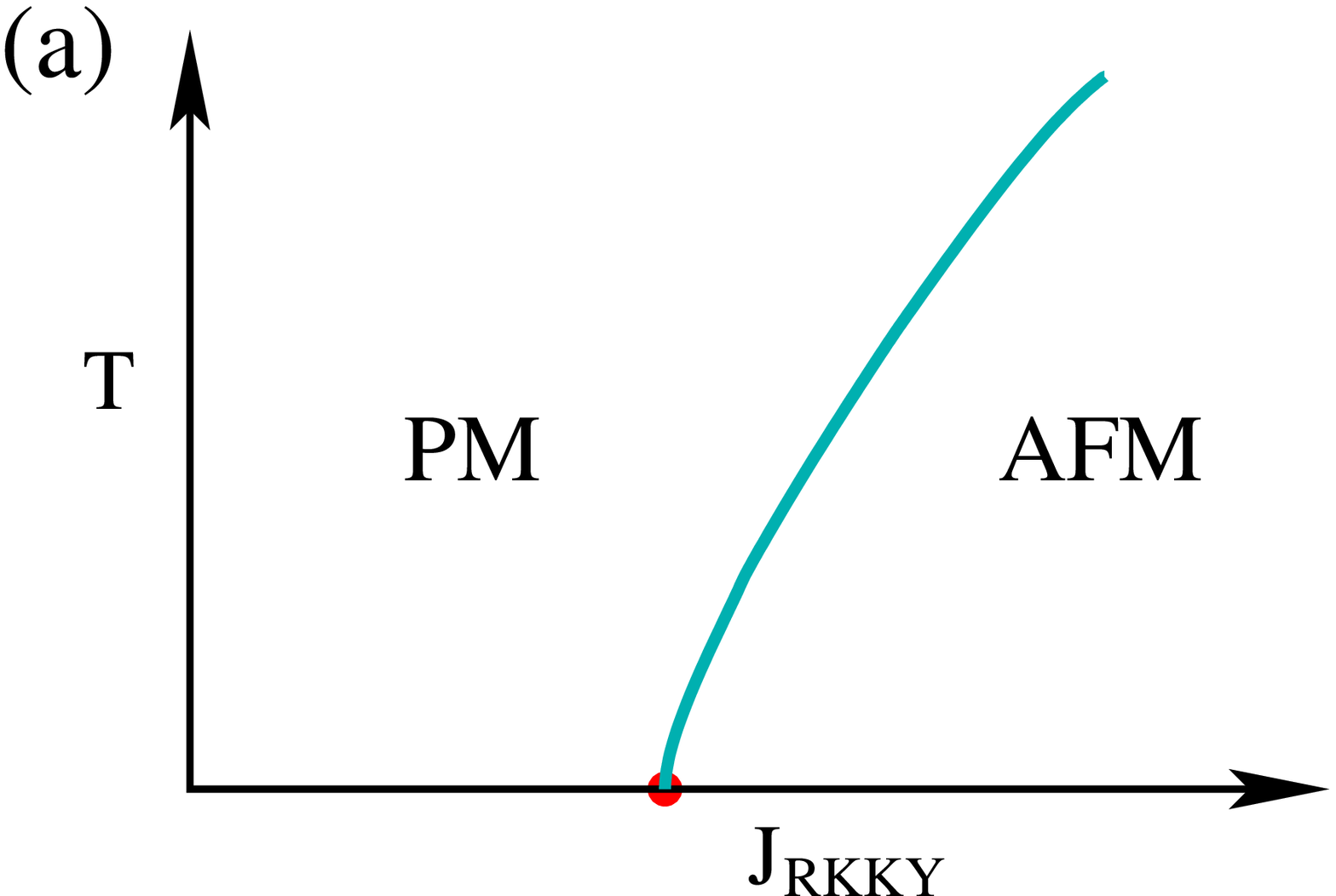}
\includegraphics[width=7.5cm]{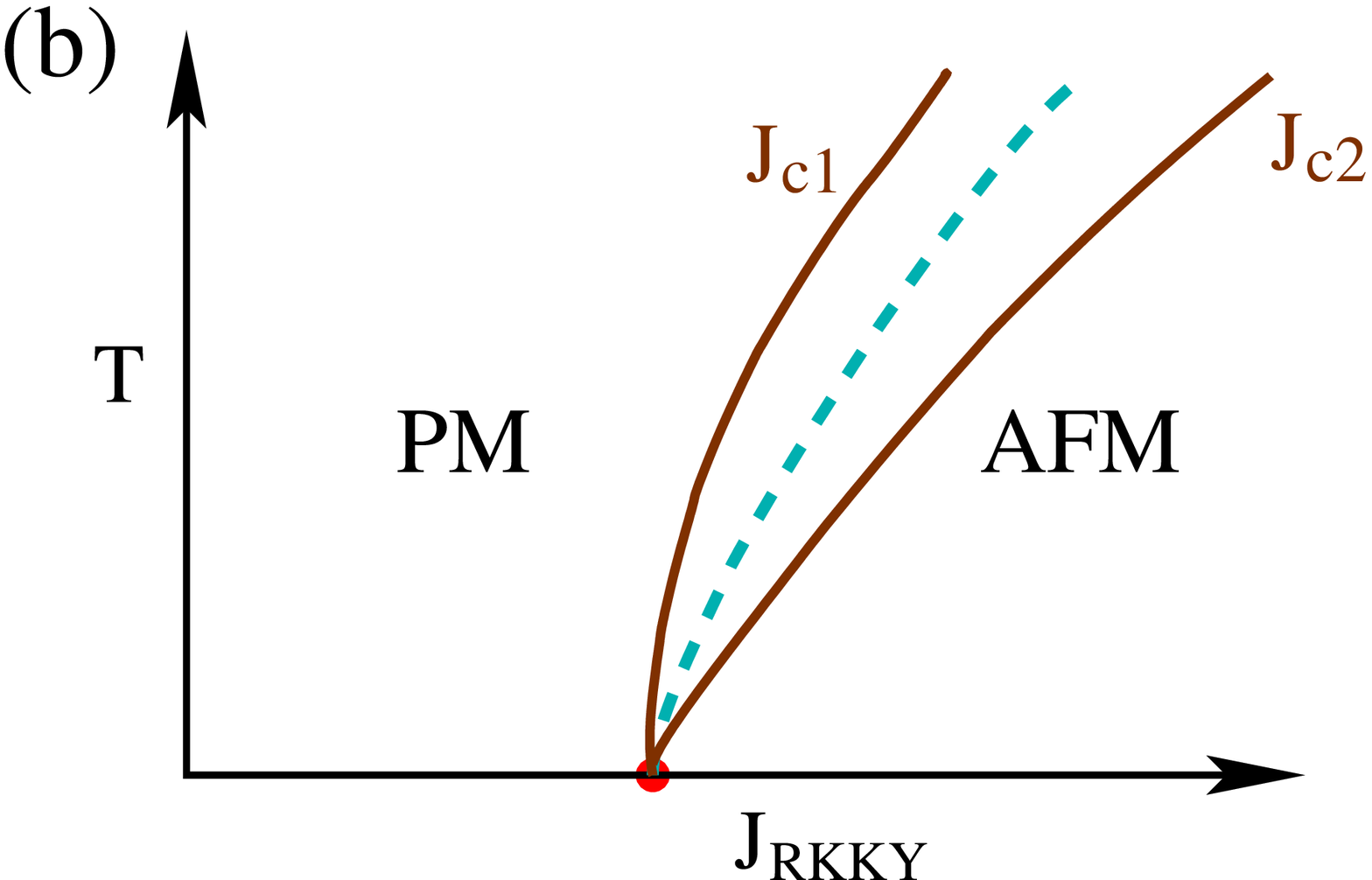}
\includegraphics[width=7.5cm]{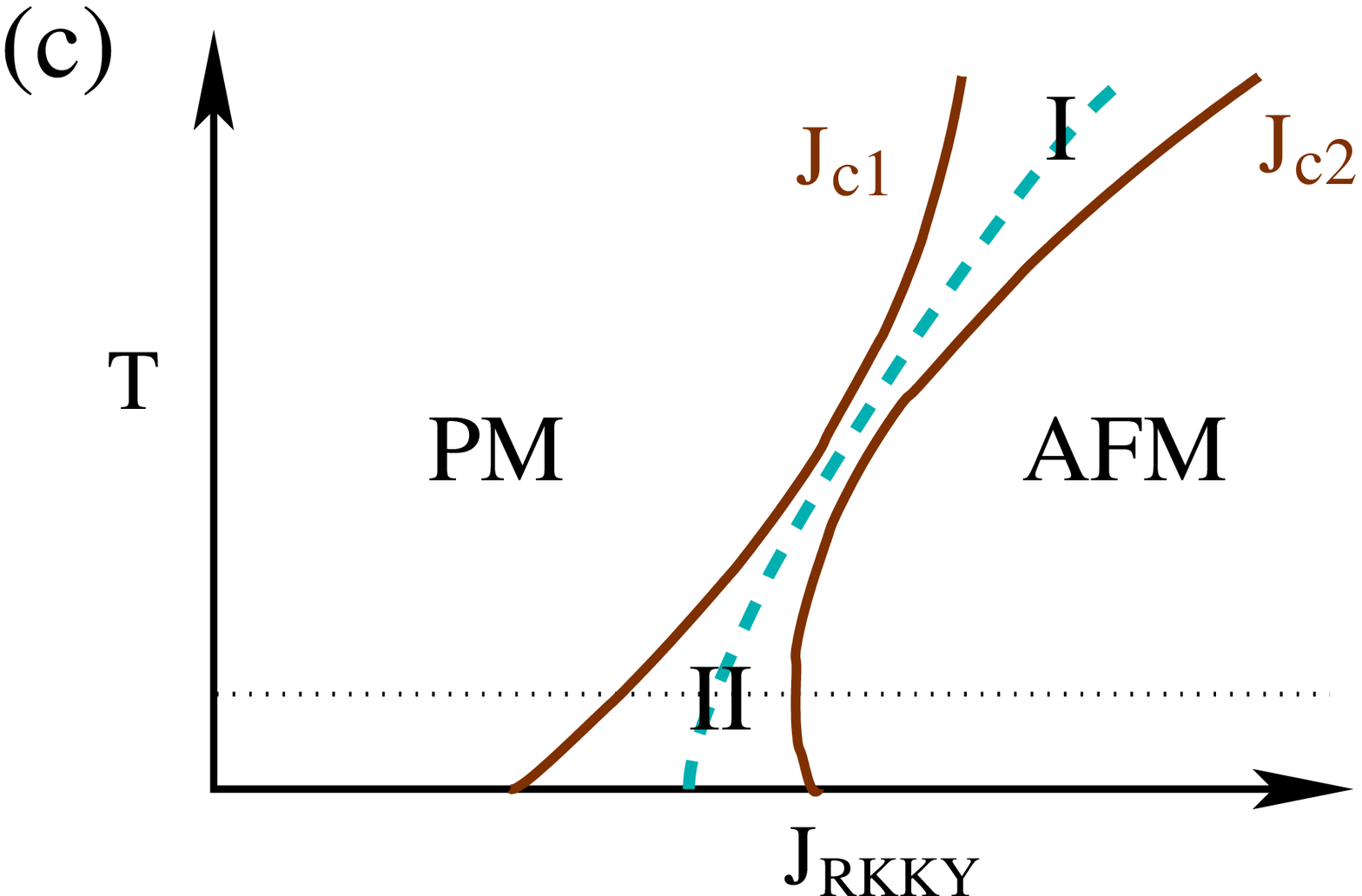}
\caption{(a) Hypothetical phase diagram of the periodic Anderson model
[see Eq.(\ref{eq-02}) below]. There is a continuous phase transition
between the PM and AFM phases and it ends up with a QCP. (b) and (c)
Two scenarios of the EDMFT phase transition, which was found to be
first order at $T>0$ \cite{si3,ping,si-note}. In the figures, the
$J_{c1}$ line is where the AFM solution disappears. At the locus there
is a finite jump in the magnetization which decreases with decreasing
temperature \cite{ping,si3}. The $J_{c2}$ line is where the spin
susceptibility at the AFM ordering wave vector diverges. In the region
in between the two phases coexist and the first order transition is
represented by a dashed curve. Panel (c) is a sketch of the results
presented in Ref.\cite{ping}. According to the calculation, the Kondo
temperature $T_K$ is near the location where the $J_{c1}$ and $J_{c2}$
lines become closest. The lowest temperature reached in
Ref.\cite{ping} is $T=0.25 T_K$, which is represented by a horizontal
dotted line in (c).}
\label{fig-01}
\end{figure}

EDMFT has been applied to study the heavy Fermions via the Kondo
\cite{si0,si1,si2,si3} and Anderson lattice \cite{ping} models. Early
EDMFT studies \cite{si0,si1,si2} approached the heavy Fermion quantum
phase transition (QPT) by following the paramagnetic (PM) solution
until where it ceased to exist and the spin susceptibility
diverged. However, the absence of the AFM phase in this scenario makes
it difficult to judge if the critical behavior is associated with a
continuous transition or the spinodal point of a first order
transition. To clarify this important issue, numerical studies of the
phase transition from both the PM and AFM sides were carried out at
finite temperatures \cite{si3,ping}. In the solution of the periodic
Anderson model (PAM) \cite{ping}, two different transitions were found
($J_{c1}$ and $J_{c2}$ lines defined in Fig.\ref{fig-01}) which
bounded a region where the PM and AFM phases coexisted. Similar
behavior was also found in the solution based on the Kondo lattice
model \cite{si3,si-note}. This strongly indicates a first order phase
transition, at least for $T>0$.\\

There are important questions, though, remain unanswered. First, given
the totally different behaviors along the mean field transition lines,
it is interesting to compare and contrast the physical meanings of the
two. Unlike at the $J_{c2}$ line where the spin susceptibility at the
AFM ordering wave vector becomes critical, it is unclear from the
EDMFT calculation itself \cite{si3,ping} which response function is
driven critical, even though critical slowing down was experienced.
Second, there are concerns with regard to a possible quantum critical
point (QCP) where the $J_{c1}$ and $J_{c2}$ lines merge [see
Fig.\ref{fig-01}(b)]. As a result, a novel quantum critical behavior
may occur. Existing analysis \cite{pankov} can not rule out such a
possibility. Besides, although our numerical results with $T \ge 0.25
T_K$ does not seem to support this scenario [see Fig.\ref{fig-01}(c)],
the temperatures reached in Ref.\cite{ping} may not be low enough to
be conclusive. The current paper is contributed to clarify these
issues, which are all related to the local spin self-energy
approximation.\\

In Sec.\ref{sec-02} we introduce the EDMFT approximation on two
sublattices via Baym-Kadanoff functional which is then used in
Sec.\ref{sec-03} to formulate the instability criteria. Technical
details of these two parts are presented in Appendices \ref{sec-a1}
and \ref{sec-a2}, respectively. Sec.\ref{sec-04} contains conclusions
and further discussions.\\

\section{EDMFT Formulation of the Periodic Anderson Model}
\label{sec-02}

\subsection{The Periodic Anderson Model}

We study the periodic Anderson model (PAM) with the local f-moments
forming a hypercubic lattice in $d$-dimensions:

\[
  H=\sum_{\vec{k}\sigma} (\epsilon_{\vec{k}}-\mu) 
  c^{\dagger}_{\vec{k}\sigma} c_{\vec{k}\sigma} 
  +V\sum_{i\sigma}(c_{i\sigma}^{\dagger}f_{i\sigma}
  +f_{i\sigma}^{\dagger}c_{i\sigma})
  +(E_f-\mu)\sum_{i\sigma}n^f_{i\sigma}
\]
\begin{equation}
\label{eq-02}
  +U\sum_{i}\left(n^f_{i\uparrow}-1/2\right)
  \left(n^f_{i\downarrow}-1/2\right)+\frac{J_{RKKY}}{d}
  \sum_{\langle ij \rangle} S^f_{i,z} S^f_{j,z}.
\end{equation}

\noindent An RKKY interaction is introduced explicitly between the
z-components of the nearest-neighboring f-electron spins,
$S^f_{i,z}=n^f_{i,\uparrow} -n^f_{i,\downarrow}$. After intergrating
out the c-electrons and introducing a Hubbard-Stratonovic field $\phi$
to decouple the interactions, we obtain:

\[
  {\cal A}=-\int_{0}^{\beta} d \tau \int_{0}^{\beta} d \tau'\sum_{ii',\sigma}
  f^{\dagger}_{\sigma}(i,\tau)
  \left[G_{0\sigma}\right]^{-1}(i\tau|i'\tau') f_{\sigma}(i',\tau')
\]
\begin{equation}
\label{eq-03}
   -\frac{1}{2}\int_0^{\beta} d\tau \int_0^{\beta} d\tau' \sum_{ii'}
   \phi(i,\tau) [D_0]^{-1}(i\tau|i'\tau') \phi(i',\tau')
   -\int_0^{\beta} d\tau \sum_{i} \phi(i,\tau) S^f_{z}(i,\tau).
\end{equation}

\noindent In this action the hybridization broadened f-band is
described by the free Green's function:

\begin{equation}
\label{eq-04}
  G_0(\vec{k},ip_n)=
  \left(ip_n+\mu-E_f-\frac{V^2}{ip_n+\mu-\epsilon_{\vec{k}}}\right)^{-1}
\end{equation}

\noindent with $p_n=(2n+1)\pi/\beta$. The free Boson Green's function
is given by:

\begin{equation}
\label{eq-05}
  D_0(\vec{k},i\omega_n)=D_0(\vec{k})=
  -U+\frac{J_{RKKY}}{d} \sum_{i=1}^{d} \cos k_i
\end{equation}

\noindent with $\omega_n=2n\pi/\beta$. Since the bare interaction is
instantaneous, the r.h.s. of Eq.(\ref{eq-05}) is frequency
independent.\\

\subsection{EDMFT via Baym-Kadanoff Formulation}

We formulate the EDMFT via Baym-Kadanoff functional \cite{chitra}:

\[
   \Gamma_{BK}[G,D,m]=
   {\rm Tr\;} \ln G - {\rm Tr\;} G_0^{-1} G
   -\frac{1}{2} {\rm Tr\;} \ln D + \frac{1}{2} {\rm Tr\;} D_0^{-1} D
\]
\begin{equation}
\label{eq-06}
   + \frac{1}{2} m D_0^{-1} m +\Phi_{EDMFT}[G_{loc},D_{loc},m],
\end{equation}

\noindent $G$ ($D$) is the full electron (Boson) Green's
function. $m=\langle \phi \rangle$. The EDMFT approximated potential
$\Phi_{EDMFT}$ is a two particle irreducible (2PI) functional of the
local Green's functions only. Since the action (\ref{eq-03}) contains
just a spatially local interaction vertex, $\Phi_{EDMFT}$ can be
written as a summation over the local contributions. On a bipartite
lattice with sublattices $A$ and $B$, this potential is given by:

\begin{equation}
\label{eq-07}
  \Phi_{EDMFT}[G_{loc},D_{loc},m]= -\int_0^{\beta} d\tau
  \sum_{j,\sigma} m(j,\tau) \sigma G_{\sigma}(j\tau^-|j\tau)
  +\sum_{j\in A} \Psi_A[G_{jj,\sigma},D_{jj}] +\sum_{l\in B}
  \Psi_B[G_{ll,\sigma},D_{ll}]
\end{equation}

\noindent where the functionals $\Psi$ contain second and higher order
diagrams in terms of the interaction vertex. To solve the AFM phase with 
the single impurity EDMFT, we need to further assume (see 
Ref.\cite{ping}),

\begin{equation}
\label{eq-08}
  \Psi_A[G_{jj,\sigma},D_{jj}]_{j\in A}=\Psi_B[G_{ll,-\sigma},D_{ll}]_{l\in B}
\end{equation}

\noindent Here the translational invariance within each sublattice is
utilized. In the PM phase the electron Green's functions are spin
independent and the assumption is still valid.\\

The Baym-Kadanoff functional gives physical solution at its stationary
point. As a result, we have,

\[
  \left[G_{\sigma}^{-1}(\vec{k},ip_n)\right]_{AB}=
  \left( \begin{array}{cc}
  ip_n+\mu-E_f & 0 \\
  0 & ip_n+\mu-E_f
  \end{array} \right)
\]
\begin{equation}
\label{eq-09}
  -\frac{V^2}{(ip_n+\mu)^2-\epsilon^2_{\vec{k}}}
  \left( \begin{array}{cc}
  ip_n+\mu & -\epsilon_{\vec{k}} \exp(+ik_x) \\
  -\epsilon_{\vec{k}} \exp(-ik_x) & ip_n+\mu \\
  \end{array} \right)
  -\left( \begin{array}{cc}
  \Sigma_{\sigma}(ip_n) & 0 \\
  0 & \Sigma_{-\sigma}(ip_n)
  \end{array} \right).
\end{equation}

\begin{equation}
\label{eq-10}
  D^{-1}(\vec{k},i\omega_n)=D_0^{-1}(\vec{k})-\Pi(i\omega_n)
\end{equation}

\begin{equation}
\label{eq-11}
  m(i,\tau)= \sum_{j,\sigma} D_{0,ij} \sigma 
  \langle f_{\sigma}^{\dagger}(j,\tau) f_{\sigma}(j,\tau) \rangle
\end{equation}

\noindent Several remarks are in place. First, due to the sublattice
structure, the electron Dyson equation (\ref{eq-09}) is in a $2\times
2$ matrix form. The electron self energy,

\begin{equation}
\label{eq-12}
  \Sigma_{X,\sigma}(ip_n) \stackrel{\rm def}{=} 
  \frac{\delta \Phi_{EDMFT}}{\delta G_{jj,\sigma}(ip_n)}|_{j\in X},
\end{equation}

\noindent with $X=A,B$, is local in space. Due to translational
invariance, we can neglect its spatial coordinates. From
Eq.(\ref{eq-08}),

\begin{equation}
\label{eq-13}
   \Sigma_{A,\sigma}(ip_n)=\Sigma_{B,-\sigma}(ip_n).
\end{equation}

\noindent As a result \cite{note-spin}, we are allowed to suppress the
sublattice index of the self-energy in Eq.(\ref{eq-09}). In the PM
phase, the self-energies are spin independent and the equation reduces
to:

\begin{equation}
\label{eq-14}
   G^{-1}(\vec{k},ip_n) = G_0^{-1}(\vec{k},ip_n)-\Sigma(ip_n).
\end{equation}

\noindent Second, in the Boson Dyson equation (\ref{eq-10}), the
self-energy is defined as:

\begin{equation}
\label{eq-15}
  \Pi_{X}(i\omega_n) \stackrel{\rm def}{=} -2 \frac{\delta
  \Phi_{EDMFT}}{\delta D_{jj}(i\omega_n)}|_{j\in X}.
\end{equation}

\noindent Eq.(\ref{eq-10}) carries a scalar form because the local
Boson self-energy is the same on both sublattices due to
symmetry. Finally, from Eq.(\ref{eq-11}), the physical order parameter
$m$ is time independent and its momentum dependence, according to
Eq.(\ref{eq-08}), is restricted to $\vec{Q}\stackrel{\rm
def}{=}(\pi,\cdots,\pi)$ for both the AFM ($m\ne 0$) and the PM
($m=0$) phases. In EDMFT, we solve the self-energies using an
effective impurity model under certain self-consistent
conditions.. [See Appendix \ref{sec-a1}]\\

\section{Instability Criteria}
\label{sec-03}

\subsection{Instability Criterion of the AFM phase ($J_{c1}$ line)}

The general instability criterion against the formation or
disappearance of a static spin density wave of wave vector $\vec{Q}$
is given by:

\begin{equation}
\label{eq-16}
   \chi^{-1}\stackrel{\rm def}{=} 
   \frac{d^2\Gamma_{BK}[G,D,m]}{dm^* dm}=0.
\end{equation}

\noindent where $m=m(\vec{Q},i0)$. Here the total derivatives are
taken on the physical manifold of the Baym-Kadanoff functional defined
through Eqs.(\ref{eq-09})-(\ref{eq-11}). As a result, the criterion
becomes, (see Appendix \ref{sec-a2} for details)

\begin{equation}
\label{eq-17}
   \chi^{-1}_{J_{c1}}\stackrel{\rm def}{=} 
    D_0^{-1}(\vec{Q})-\int_0^{\beta} d\tau
   \sum_{a,b=1}^{2} (-1)^{a+b}
   \Pi_{J_{c1},\vec{Q}}[(\tau|\tau),(0|0)]=0
\end{equation}

\noindent where

\begin{equation}
\label{eq-18}
  \bigglb[\Pi_{J_{c1},\vec{Q}}\biggrb]^{-1}
  [(\tau_1|\tau_1'),(\tau_2|\tau_2')]
  =\bigglb[ \chi_{0,\vec{Q}}^{-1}-\chi_{0,imp}^{-1}+
  \chi_{imp}^{-1}+\tilde{\cal D}_0 \biggrb] 
  [(\tau_1|\tau_1'),(\tau_2|\tau_2')]
\end{equation}

\noindent In the above equation $\chi_{0,\vec{Q}}$ is an electron-hole
bubble evaluated with the full Green's functions at the wave vector
$\vec{Q}$. $\chi_{0,imp}$ is a smiliar bubble obtained via the full
impurity Green's function. $\chi_{imp}$ is a four point response
function of the impurity model. $\tilde{\cal D}_0$, which depends on
${\cal D}_0$, is the bare interaction in the impurity
model. Eq.(\ref{eq-17}) gives the general EDMFT instability criterion
without further approximation and applies to the $J_{c1}$ line where
the AFM solution at $\vec{Q}=(\pi,\cdots,\pi)$ disappears. The
existence of the electron-hole bubble at the ordering wave vector in
Eq.(\ref{eq-18}) reveals the fact that even in the infinite
coordination limit where the mean field method becomes exact, there is
still a non-vanishing momentum-dependent contribution in the effective
spin susceptibility. The matrix inversions in Eq.(\ref{eq-18}) involve
matrices labeled by four time coordinates, two for the row and two for
the column, respectively. As a result, this expression is generally
very complicated and can not be further simplified.\\

\subsection{Instability Criterion of the PM phase ($J_{c2}$ line)}

In the EDMFT of the PM phase, the effective susceptibility given in
Eq.(\ref{eq-18}) contributes directly to the spin self-energy
\cite{pankov} and, as a result, should be local in space. This means
we need further to restrict $\chi_{0}$ to be local. However, from the
EDMFT self-consistency that the local Green's functions on the lattice
equal to the impurity ones, $\chi_{0,loc}$=$\chi_{0,imp}$. Hence the
instability criterion becomes

\begin{equation}
\label{eq-19}
   \chi^{-1}_{J_{c2}}\stackrel{\rm def}{=} 
   D_0^{-1}(\vec{Q})-\int_0^{\beta} d\tau
   \sum_{a,b=1}^{2} (-1)^{a+b}
   \Pi_{J_{c2}}[(\tau|\tau),(0|0)]=0
\end{equation}

\noindent where

\begin{equation}
\label{eq-20}
  \bigglb[\Pi_{J_{c2}}\biggrb]^{-1}
  [(\tau_1|\tau_1'),(\tau_2|\tau_2')]
  =\bigglb[ \chi_{imp}^{-1}+\tilde{\cal D}_0 \biggrb] 
  [(\tau_1|\tau_1'),(\tau_2|\tau_2')]
\end{equation}

\noindent The special form of the matrix ${\cal D}_0$ [see
Eq.(\ref{eq-a-10})] allows us to carry out the matrix operations
explicitly and obtain,

\begin{equation}
\label{eq-21}
   D_0(\vec{Q})={\cal D}_0(i0)+\chi_{zz}^{-1}(i0)=\Pi^{-1}(i0)
\end{equation}

\noindent with 

\begin{equation}
\label{eq-22}
  \chi_{zz}(\tau) \stackrel{\rm def}{=} \langle S_z^f(\tau)S_z^f(0)\rangle
   \equiv [\chi_{imp,G_{\uparrow
   }G_{\uparrow }}-\chi_{imp,G_{\uparrow
   }G_{\downarrow}}-\chi_{imp,G_{\downarrow}G_{\uparrow }}
   +\chi_{imp,G_{\downarrow}G_{\downarrow}}][(\tau|\tau),(0|0)]
\end{equation}

\noindent The last equality in Eq.(\ref{eq-21}) is derived by an
identity \cite{identity}. Comparing with the Boson Dyson equation
(\ref{eq-10}), we see that the above instability criterion is
identical to the Stoner criterion in which the divergence of the
magnetic susceptibility at $\vec{Q}=(\pi,\cdots,\pi)$ signals phase
transition \cite{note}.\\

\section{Conclusions and Discussions}
\label{sec-04}

We have derived in this paper the phase instability criterion of the
EDMFT solution to the periodic Anderson model for both the AFM and PM
phases. The generic instability criterion (which applies to $J_{c1}$
line in Fig.\ref{fig-01}) involves an effective spin susceptibility,
Eq.(\ref{eq-17}), and is different from that used to determine the
transition line ($J_{c2}$) bounding the PM phase,
Eq.(\ref{eq-19}). The difference is in an extra electron-hole bubble
at the AFM ordering wave-vector in the former. This bubble is momentum
dependent and survives in the infinite coordination limit. As a
result, at the locus where one of the phases reaches the instability
condition, the other one remains stable. This explains the phase
coexistence. It persists to $T=0$ since the electron-hole bubble
remains non-zero. This is consistent with what we obtained numerically
\cite{ping} in Region II of Fig.\ref{fig-01}(c). We should point out,
though, at dimensions $d>4$ and temperatures $T \gtrsim T_{Kondo}$
[Region I in Fig.\ref{fig-01}(c)], the difference between the $J_{c1}$
and $J_{c2}$ lines becomes negligibly small \cite{ping1}. This is due
to the spatial correlation becoming weaker at higher dimensions
\cite{pankov} and temperatures. We note in passing that no matter
which criterion is satisfied, the divergence of the corresponding
effective spin susceptibility at the AFM ordering wave vector
naturally results in the divergence of the local spin susceptibility
as long as the spin fluctuations are two dimensional
\cite{si0,ping}. This is a result of the dimensionality and has
nothing to do with the spin self-energy being local in space.\\

The true mean field transition is thus first order and lies between
the $J_{c1}$ and $J_{c2}$ lines where the free energies of the two
phases cross. Physically, the two sublattice EDMFT (as applied in the
AFM phase) contains in its instability criterion an electron-hole
bubble, which serves as a rough description of the feedback from the
electron-hole excitations to the spin response. However, this feedback
does not appear explicitly in the EDMFT self-consistency, which is
evident from what we described in Appendix \ref{sec-a1}. As a result,
the EDMFT spin susceptibility, which is different from the physical
one in the instability criterion, Eq.(\ref{eq-17}), does not
experience any singularity as the $J_{c1}$ line is crossed. On the
other hand, the homogeneous EDMFT (as applied in the PM phase),
contains the same singular behavior in the spin response as that in
the instability criterion, Eq.(\ref{eq-19}). As a result, when the
phase boundary is approached, EDMFT is able to adjust
self-consistently to reflect the singular behavior in the spin
channel. However, as we have already noted, the problem on this side
is that the feedback from the non-local electron-hole excitations is
totally missing. So both the transition lines contain unphysical
features, and neither of them, as far as the critical properties are
concerned, is close to the true transition. A related issue, which
concerns the critical exponent $\alpha$ in Eq.(\ref{eq-01}) along the
$J_{c2}$ line, further supports our conclusion. It was shown that at
$T=0$ on the $J_{c2}$ line, the critical frequency dependence could
not develop a sublinear form \cite{pankov1}.\\

After all, it is not a surprise that, although it works well
qualitatively in describing many other physical properties
\cite{ping}, the EDMFT fails to capture the right phase
transition. This is certainly one of the issues one needs to improve
over the mean field approach. Given what we have concluded in this
paper, it seems important that one needs to find a way allowing proper
feedback from the electron-hole excitations, which is spatially
non-local, to the f-electron spin response. A natural way to proceed
is to combine the EDMFT scheme with the random phase approximation
(RPA) \cite{ping2}. In this combination, the spin self-energy contains
the local EDMFT part together with the non-local RPA part. This is a
desirable feature as one can see from the EDMFT instability criterion
Eq.(\ref{eq-18}). Besides, the scheme is derivable from the
Baym-Kadanoff functional \cite{ping2}. Of course, with the new scheme,
the instability criterion itself is modified and its implication to
the heavy Fermion phase transition has not yet been explored. A
different route is to utilize the cellular DMFT \cite{cdmft}. To this
end, a two impurity Anderson model subject to the DMFT self-consistent
electron bath results in a qualitative improvement \cite{ping3}. In
this formalism, the RKKY interaction is generated dynamically, instead
of being added in by hand as in Eq.(\ref{eq-02}). The spin
susceptibility across the two impurity sites, which contains the
corresponding electron-hole bubble as the lead order contribution,
renders a limited momentum dependence and turns out to be essential to
the improvement.\\

\acknowledgements

The authors would like to acknowledge helpful discussions with
E. Abrahams, P. Coleman, A. Georges, S. Pankov, and A. Schiller. This
research was supported by NSF under No. DMR-0096462 and the Center for
Materials Theory at Rutgers University.\\

\appendix

\section{Effective Impurity Model and EDMFT Self-Consistency}
\label{sec-a1}

To obtain the local self-energies, we need to solve an effective
impurity model:

\[
  {\cal A}_{0}^{eff}=-\int_{0}^{\beta} d\tau \int_{0}^{\beta} d\tau'
  \sum_{\sigma} f^{\dagger}_{\sigma}(0,\tau)
  \left[{\cal G}_{0\sigma}\right]^{-1}(\tau-\tau') f_{\sigma}(0,\tau')
\]
\begin{equation}
\label{eq-a1-01}
   -\frac{1}{2}\int_0^{\beta} d\tau \int_0^{\beta} d\tau'
   \phi(0,\tau) {\cal D}_0^{-1}(\tau-\tau') \phi(0,\tau')
   -\int_0^{\beta} d\tau \phi(0,\tau) S^f_{z}(0,\tau).
\end{equation}

\noindent The mean field Weiss functions ${\cal G}_{0\sigma}$ and
${\cal D}_0$ are decided by the following self-consistent conditions:

\begin{equation}
\label{eq-a1-02}
  {\cal G}^{-1}_{0\sigma}(ip_n)=[\sum_{\vec{k}} 
  G_{\sigma}(k,ip_n)]^{-1}+\Sigma^{imp}_{\sigma}(ip_n)
\end{equation}
\begin{equation}
\label{eq-a1-03}
  {\cal D}^{-1}_{0}(i\omega_n)=[\sum_{\vec{k}}
  D_{\sigma}(\vec{k},i\omega_n)]^{-1}+\Pi^{imp}(i\omega_n)
\end{equation}

\noindent with $G_{\sigma}(\vec{k},ip_n)$ and
$D_{\sigma}(\vec{k},i\omega_n)$ given by Eqs.(\ref{eq-09}) and
(\ref{eq-10}), respectively. The self-energies are,

\begin{equation}
\label{eq-a1-04}
   \Sigma^{imp}_{\sigma}(ip_n)=[{\cal G}_{0\sigma}]^{-1}(ip_n)-
   [G^{imp}_{\sigma}]^{-1}(ip_n)
\end{equation}
\begin{equation}
\label{eq-a1-05}
   \Pi^{imp}(i\omega_n)=[{\cal D}_{0}]^{-1}(i\omega_n)-
   [D^{imp}]^{-1}(i\omega_n)
\end{equation}

\noindent where the impurity Green's functions $G^{imp}_{\sigma}$ and
$D^{imp}$ are obtained by solving the effective action
(\ref{eq-a1-01}). In Eqs.(\ref{eq-09}) and (\ref{eq-10}), we need to
use the lattice self-energies which are usually assumed to be the same
as the impurity ones in the disordered phase. In the ordered phase,
the electron self-energy on the lattice is different from that of the
impurity model by a Hartree term, while the Boson self-energy is still
the same,

\begin{equation}
\label{eq-a1-06}
   \Sigma_{\sigma}(ip_n)=\Sigma^{imp}_{\sigma}(ip_n)
   -\sigma [{\cal D}_0(i0)-(U-J_{RKKY})] \langle S_z^f \rangle
\end{equation}
\begin{equation}
\label{eq-a1-07}
   \Pi(i\omega_n)=\Pi^{imp}(i\omega_n)
\end{equation}

\noindent The meaning of Eq.(\ref{eq-a1-06}) is that we need to
replace the Hartree self-energy of the impurity model by that on the
lattice, using the electron magnetization. This procedure is related
to Eq.(\ref{eq-11}), which would otherwise introduce a third
self-consistent equation. With this, we have presented a complete
self-consistent loop.\\

\section{Derivation of the Instability Criterion for the AFM phase}
\label{sec-a2}

We derive here the instability criterion specific to the periodic
Anderson model. From the general condition, Eq.(\ref{eq-16}), together
with Eqs.(\ref{eq-09})-(\ref{eq-11}), we obtain,

\begin{eqnarray}
\label{eq-a-01}
   \frac{\partial^2\Gamma_{BK}}{\partial m^* \partial m} 
   + \int dx \frac{\partial G_{\sigma_x}(x)}{\partial m^*}
   \frac{\partial^2\Gamma_{BK}}{\partial G_{\sigma_x}(x) \partial m}
   + \int dx \frac{\partial D(x)}{\partial m^*}
   \frac{\partial^2\Gamma_{BK}}{\partial D(x) \partial m} =0 \\
\label{eq-a-02}
   \int dx \frac{\partial G_{\sigma_x}(x) }{\partial m^*}
   \frac{\partial^2\Gamma_{BK}}{\partial G_{\sigma_x}(x)\partial
     G_{\sigma_y}(y)}
   + \int dx \frac{\partial D(x) }{\partial m^*}
   \frac{\partial^2\Gamma_{BK}}{\partial D(x) \partial G_{\sigma_y}(y)}
   + \frac{\partial^2\Gamma_{BK}}{\partial m^* \partial G_{\sigma_y}(y)} =0 \\
\label{eq-a-03}
   \int dx \frac{\partial G_{\sigma_x}(x) }{\partial m^*}
   \frac{\partial^2\Gamma_{BK}}{\partial G_{\sigma_x}(x) \partial D(y)}
   + \int dx \frac{\partial D(x) }{\partial m^*}
   \frac{\partial^2\Gamma_{BK}}{\partial D(x) \partial D(y)}
   + \frac{\partial^2\Gamma_{BK}}{\partial m^* \partial D(y)} =0.
\end{eqnarray}

\noindent We used $x=(\vec{R}_j,\tau|\vec{R}_{j'},\tau')$, (similar
for $y$) and $\int dx = \sum_{j,j'} \int_0^{\beta} d\tau
\int_0^{\beta} d\tau'$. Summation over the repeated spin indices is
implied. Solving $\partial G/\partial m^*$ and $\partial D/\partial
m^*$ from Eqs.(\ref{eq-a-02}) and (\ref{eq-a-03}), and substituting
them in Eq.(\ref{eq-a-01}), we obtain:

\[
  \frac{\partial^2\Gamma_{BK}[G,D,m]}{\partial m^* \partial m} -\int
  dx \int dy \left[\partial^2\Gamma_{BK}/\partial m^* \partial G_{\sigma_x}(x),
  \partial^2\Gamma_{BK}/\partial m^* \partial D(x)\right]
\]
\begin{equation}
\label{eq-a-04}
  \times\left[ \begin{array}{cc}
   \partial^2 \Gamma_{BK} / \partial G_{\sigma_x} \partial G_{\sigma_y}  &
   \partial^2 \Gamma_{BK} / \partial G_{\sigma_x} \partial D  \\
   \partial^2 \Gamma_{BK} / \partial D \partial G_{\sigma_y}  &
   \partial^2 \Gamma_{BK} / \partial D \partial D 
   \end{array} \right]^{-1}(x,y)
   \left[ \begin{array}{c}
     \partial^2\Gamma_{BK}/ \partial G_{\sigma_y}(y) \partial m\\
     \partial^2\Gamma_{BK}/ \partial D(y) \partial m
           \end{array} \right]=0.
\end{equation}

\noindent Using Eqs.(\ref{eq-06}) and (\ref{eq-07}), we find
$\partial^2\Gamma_{BK}/ \partial G_{\sigma}(j\tau_1|j'\tau_1^{'})
\partial m(i,\tau) = -\sigma \delta_{ij} \delta_{ij'}
\delta(\tau-\tau_1)\delta(\tau-\tau_1^{'}) $ and $
\partial^2\Gamma_{BK}/ \partial D(j\tau_1|j'\tau_1^{'}) \partial
m(i,\tau) = 0$. Besides, $ \partial^2\Gamma_{BK}[G,D,m]/\partial m^*
\partial m= D_0^{-1}(\vec{Q})$. So we have:

\begin{equation}
\label{eq-a-05}
   \beta D_0^{-1}(\vec{Q}) =\frac{1}{N}\sum_{j_1}\int_0^{\beta} d\tau_1
   \sum_{j_2}\int_0^{\beta} d\tau_2 \exp(-i\vec{Q}\cdot\vec{R}_{j_1})
   \exp( i\vec{Q}\cdot\vec{R}_{j_2})
\end{equation}
\[
   \times
   (1,-1,0)
   \left[ \begin{array}{ccc}
   \Gamma^{(2)}_{G_{\uparrow}  G_{\uparrow}  } &
   \Gamma^{(2)}_{G_{\uparrow}  G_{\downarrow}} &
   \Gamma^{(2)}_{G_{\uparrow}  D             } \\
   \Gamma^{(2)}_{G_{\downarrow}G_{\uparrow}  } &
   \Gamma^{(2)}_{G_{\downarrow}G_{\downarrow}} &
   \Gamma^{(2)}_{G_{\downarrow}D             } \\
   \Gamma^{(2)}_{D             G_{\uparrow}  } &
   \Gamma^{(2)}_{D             G_{\downarrow}} &
   \Gamma^{(2)}_{D             D             } 
   \end{array} \right]^{-1} [(j_1\tau_1|j_1\tau_1),(j_2\tau_2|j_2\tau_2)]
   \left( \begin{array}{r}
   1\\-1\\0 \end{array} \right),
\]

\noindent with $ \Gamma^{(2)}_{XY}
[(j_1\tau_1|j_1'\tau_1'),(j_2\tau_2|j_2'\tau_2')] =
\partial^2\Gamma_{BK}/\partial X(j_1\tau_1|j_1'\tau_1') \partial
Y(j_2\tau_2|j_2'\tau_2')$ for $X,Y=G_{\sigma},D$. To solve the matrix
$\Gamma^{(2)}$, we use again the Baym-Kadanoff functional
(\ref{eq-06}) and obtain:

\begin{equation}
\label{eq-a-06}
  \Gamma^{(2)}_{XY} [(j_1\tau_1|j_1'\tau_1'),(j_2\tau_2|j_2'\tau_2')]
  = \chi_{0,XY}^{-1}[(j_1\tau_1|j_1'\tau_1'),(j_2\tau_2|j_2'\tau_2')]
  + \Phi^{(2)}_{XY} [(j_1\tau_1|j_1'\tau_1'),(j_2\tau_2|j_2'\tau_2')]
\end{equation}

\noindent where 

\begin{equation}
\label{eq-a-07}
  \chi_{0,XY}[(j_1\tau_1|j_1'\tau_1'),(j_2\tau_2|j_2'\tau_2')]
  \stackrel{\rm def}{=}
  \left\{
  \begin{array}{cc}
  -G_{\sigma_{X'},\sigma_Y}(j_1'\tau_1'|j_2\tau_2) 
   G_{\sigma_{Y'},\sigma_X}(j_2'\tau_2'|j_1\tau_1), & X,Y=G_{\sigma}\\
   D(j_1\tau_1|j_2\tau_2)D(j_1'\tau_1'|j_2'\tau_2')
  +D(j_1\tau_1|j_2'\tau_2')D(j_1'\tau_1'|j_2\tau_2),
  \hspace*{0.2cm} & X,Y=D \\
   0,& {\rm else}
  \end{array}\right.
\end{equation}

\begin{equation}
\label{eq-a-08}
  \Phi^{(2)}_{XY} [(j_1\tau_1|j_1'\tau_1'),(j_2\tau_2|j_2'\tau_2')]
  \stackrel{\rm def}{=}
  \frac{\partial^2\Phi_{EDMFT}}{ \partial X(j_1\tau_1|j_1'\tau_1') \partial
  Y(j_2\tau_2|j_2'\tau_2')}
\end{equation}

\noindent $\Phi^{(2)}_{XY}$, same as $\Phi_{EDMFT}$, contains only
propagators local in space and is 2PI in separating the external legs
labeled by 1 and 1' from those 2 and 2'. It follows then
\cite{pankov},

\begin{equation}
\label{eq-a-09}
  \bigglb[\Phi^{(2)}\biggrb]_{XY}
  [(j_1\tau_1|j_1'\tau_1'),(j_2\tau_2|j_2'\tau_2')]
  =\delta_{j_1,j_1'} \delta_{j_1,j_2} \delta_{j_1,j_2'} 
  \bigglb[-\chi_{0,imp}^{-1}+\chi_{imp}^{-1}+\tilde{\cal D}_0\biggrb]_{XY}
  [(\tau_1|\tau_1'),(\tau_2|\tau_2')].
\end{equation}

\noindent Here $\chi_{0,imp}$ is similar to that defined in
Eq.(\ref{eq-a-07}) except being local in space. We also defined:

\begin{equation}
\label{eq-a-10}
  \tilde{\cal D}_0 [(\tau_1|\tau_1'),(\tau_2|\tau_2')]
  \stackrel{\rm def}{=} \delta(\tau_1-\tau_1') \delta(\tau_2-\tau_2')
  \left[ \begin{array}{ccc}
  \;\; {\cal D}_{0}(\tau_1-\tau_2) &   -{\cal D}_{0}(\tau_1-\tau_2) &  0 \\
  -{\cal D}_{0}(\tau_1-\tau_2) &    \;\;{\cal D}_{0}(\tau_1-\tau_2) &  0 \\
    0   &   0    &  0
  \end{array} \right]
\end{equation}

\begin{equation}
\label{eq-a-11}
  \chi_{imp,XY} [(0\tau_1|0\tau_1'),(0\tau_2|0\tau_2')]
  \stackrel{\rm def}{=}
  \langle T_{\tau} \; :\hat{\cal O}_X^{\dagger}(0,\tau_1) 
                       \hat{\cal O}_X(0,\tau_1'):
  :\hat{\cal O}_Y^{\dagger}(0,\tau_2) 
   \hat{\cal O}_Y^{\dagger}(0,\tau_2'): \rangle
\end{equation}

\noindent where $\hat{\cal O}_X=c_{\sigma}$ ($\phi$) if $X=G_{\sigma}$
($D$). The instability criterion becomes:

\[
   D_0^{-1}(\vec{Q})=\frac{1}{N}\sum_{j_1,j_2}\int_0^{\beta} d\tau
   \exp(-i\vec{Q}\cdot\vec{R}_{j_1})
   \exp( i\vec{Q}\cdot\vec{R}_{j_2})
\]
\begin{equation}
\label{eq-a-12}
   \sum_{a,b=1}^{2} (-1)^{a+b}
   \left[\chi_{0}^{-1}-\chi_{0,imp}^{-1}+\chi_{imp}^{-1}+\tilde{\cal
   D}_0 \right]_{a,b}^{-1}[(j_1\tau|j_1\tau),(j_20|j_20)].
\end{equation}

\noindent where all the four terms in the square parenthesis are
$3\times 3$ matrices and after matrix inversion, only the first
$2\times 2$ block contributes. It should be noted that the matrices
are also labeled by the two pairs of the space-time coordinates and
any matrix operation should take these into account. As a result, {\it
e.g.}, in Eq.(\ref{eq-a-12}) the full matrix, labeled by
$[(j_1\tau_1|j_1'\tau_1'),(j_2\tau_2|j_2'\tau_2')]$, should be
inverted first and only after that, we set the labels to be
$[(j_1\tau|j_1\tau),(j_20|j_20)]$.\\

Finally, since in Eq.(\ref{eq-a-12}), $\chi_{0}$ is the only term
contains spatially non-local contributions, the Fourier transform
over the lattice coordinate can be taken into the matrix inversion,
which gives:

\begin{equation}
\label{eq-a-13}
   D_0^{-1}(\vec{Q})=\int_0^{\beta} d\tau
   \sum_{a,b=1}^{2} (-1)^{a+b}
   \chi_{J_{c1},\vec{Q}}[(\tau|\tau),(0|0)]
\end{equation}

\noindent where

\begin{equation}
\label{eq-a-14}
  \bigglb[\chi_{J_{c1},\vec{Q}}\biggrb]^{-1}
  [(\tau_1|\tau_1'),(\tau_2|\tau_2')]
  =\bigglb[ \chi_{0,\vec{Q}}^{-1}-\chi_{0,imp}^{-1}+
  \chi_{imp}^{-1}+\tilde{\cal D}_0 \biggrb] 
  [(\tau_1|\tau_1'),(\tau_2|\tau_2')]
\end{equation}

\noindent This gives an instability criterion consistent with the
EDMFT Baym-Kadanoff functional, Eqs.(\ref{eq-06}) and (\ref{eq-07}),
without any further approximation.\\

As it turns out, we need further to assume $\chi_{0}$ be spatially
local in Eq.(\ref{eq-a-12}), in order to describe the PM phase. In
such a case, $\chi_{0} \rightarrow \chi_{0,loc}$, we have (1) the
momentum dependent phase factors in Eq.(\ref{eq-a-12}) cancel out and
(2) $\chi_{0,imp}$ cancels $\chi_{0,loc}$ due to the EDMFT
self-consistency that the local lattice Green's functions equal to the
impurity ones.\\


\begin{thebibliography}{99}
\bibitem[17]{note}

  To see this point, one needs to remove ``$-U$'' from both $D_0$ and
  $\Pi^{-1}$ \cite{identity}. Since $\Pi(i0)<0$, it follows that the
  instability is first seen when $\vec{Q}=(\pi,\cdots,\pi)$.
\end{thebibliography}
\end{document}